\documentclass[aps,showdoi,showkeys,superscriptaddress,reprint]{revtex4-1}

\usepackage{ae,lmodern}
\usepackage[english]{babel}
\usepackage[latin1]{inputenc}
\usepackage[T1]{fontenc}
\usepackage{textcomp}
\usepackage[euler]{textgreek}
\usepackage{url}

\usepackage{amsfonts}
\usepackage{amsmath}
\usepackage{amssymb}
\usepackage{graphicx}
\usepackage{bm}
\usepackage{xcolor}

\usepackage{esvect}
\usepackage{wasysym}

\usepackage{chemist}

\usepackage{array,multirow,makecell,hhline}
\usepackage{tabularx}
\setcellgapes{1pt}
\makegapedcells
\newcolumntype{R}[1]{>{\raggedleft\arraybackslash }b{#1}}
\newcolumntype{L}[1]{>{\raggedright\arraybackslash }b{#1}}
\newcolumntype{C}[1]{>{\centering\arraybackslash }b{#1}}

\begin{document}

\title{Azimuthal instability of the radial thermocapillary flow around a hot bead trapped at the water\,--\,air interface}
\author{G. Koleski}
\affiliation{Univ. Bordeaux, CNRS, Laboratoire Ondes et Mati\`ere d'Aquitaine (UMR 5798), F-33400 Talence, France}
\affiliation{Univ. Bordeaux, CNRS, Centre de Recherche Paul Pascal (UMR 5031), F-33600 Pessac, France}
\author{A. Vilquin}
\affiliation{ESPCI, CNRS, Institut Pierre\,--\,Gilles de Gennes, Laboratoire Gulliver (UMR 7083), F-75005 Paris, France}
\author{J.-C. Loudet}
\affiliation{Univ. Bordeaux, CNRS, Centre de Recherche Paul Pascal (UMR 5031), F-33600 Pessac, France}
\affiliation{Department of Mathematics, University of British Columbia, Vancouver BC V6T 1Z2, Canada}
\author{T. Bickel}
\affiliation{Univ. Bordeaux, CNRS, Laboratoire Ondes et Mati\`ere d'Aquitaine (UMR 5798), F-33400 Talence, France}
\author{B. Pouligny}
\affiliation{Univ. Bordeaux, CNRS, Centre de Recherche Paul Pascal (UMR 5031), F-33600 Pessac, France}

\begin{abstract}
We investigate the radial thermocapillary flow driven by a laser-heated microbead in partial wetting at the water-air interface. Particular attention is paid to the evolution of the convective flow patterns surrounding the hot sphere as the latter is increasingly heated. The flow morphology is nearly axisymmetric at low laser power ($\mathcal{P}$). Increasing $\mathcal{P}$ leads to symmetry breaking with the onset of counter-rotating vortex pairs. The boundary condition at the interface, close to no-slip in the low-$\mathcal{P}$ regime, turns about stress-free between the vortex pairs in the high-$\mathcal{P}$ regime. These observations strongly support the view that surface-active impurities are inevitably adsorbed on the water surface where they form an elastic layer. The onset of vortex pairs is the signature of a hydrodynamic instability in the layer response to the centrifugal forced flow. Interestingly, our study paves the way for the design of active colloids able to achieve high-speed self-propulsion \textit{via} vortex pair generation at a liquid interface.
\end{abstract}


\maketitle

\section{Introduction}
\label{Introduction}

Temperature gradients arising along a fluid interface are responsible for local variations of its surface tension. Shear stresses ensue and set the liquid into motion. This phenomenon is known as the Marangoni effect~\cite{Scriven1960} and the resulting flow is said thermocapillary. Such flows are ubiquitous in everyday life~\cite{Fournier1992}\,--\,\cite{Keiser2017}. Over the past few decades, thermocapillary convection has attracted great interest in many research areas.

As an example, Marangoni\,--\,driven instabilities have been recognized to severely damage the quality of crystals grown from the melt~\cite{Schwabe1988, Kamotani1999} as well as that of arc welds~\cite{Oreper1983}. Interested readers can refer to reviews~\cite{Davis1987, SchatzAndNeitzel2001} for a detailed presentation of thermocapillary instabilities.

Surface forces dominate over body forces at small scales due to large surface\,--\,to\,--\,volume ratios. Surface tension driven flows thus appear as ideal candidates for the manipulation of microfluidic systems. For instance, many efforts have been devoted in recent years to the thermocapillary actuation of droplets sitting on the free surface of immiscible liquid films~\cite{YakhsiTafti2010, Rybalko2004}. It has also been shown that breaking the symmetry of recirculation flows within a microfluidic droplet allows for efficient chaotic mixing~\cite{Cordero2009}. In Ref.~\cite{Miniewicz2017}, gas bubbles forming on a thin light\,--\,absorbing liquid layer were remotely trapped and manipulated using an optical tweezer. The authors of~\cite{Namura2015} tailored laser\,--\,induced vortical flow patterns around a microbubble to sort polystyrene beads by size.

Researchers are now going further and further in the design of lab\,--\,on\,--\,a\,--\,chip platforms whose processing units enable combined and automatized operations on microfluidic systems such as droplet production, size\,--\,selective sorting, pumping, division and fusion, transport along microchannels or trapping~\cite{Baroud2007}\,--\,\cite{Basu2007}. Ref.~\cite{Davanlou2015} is a typical example in which heaters embedded on a printed circuit board are engineered to make levitated droplets collide and merge their reagents \textit{via} thermocoalescence. A holistic view of thermocapillary\,--\,based microfluidics is provided in the exhaustive reviews~\cite{Karbalaei2016, Darhuber2005}.

Light\,--\,to\,--\,work conversion has also been harnessed in the field of active matter. Recently, Maggi \textit{et al.} have shown how to rotate asymmetrically\,--\,toothed microgears sitting on a liquid\,--\,air interface by coating them on one side with a light\,--\,absorbing material~\cite{Maggi2015}. Their work is one more example that promotes thermocapillarity as a highly efficient mechanism for colloidal self\,--\,propulsion.

\begin{figure}
\centering
\includegraphics[width=0.5\columnwidth]{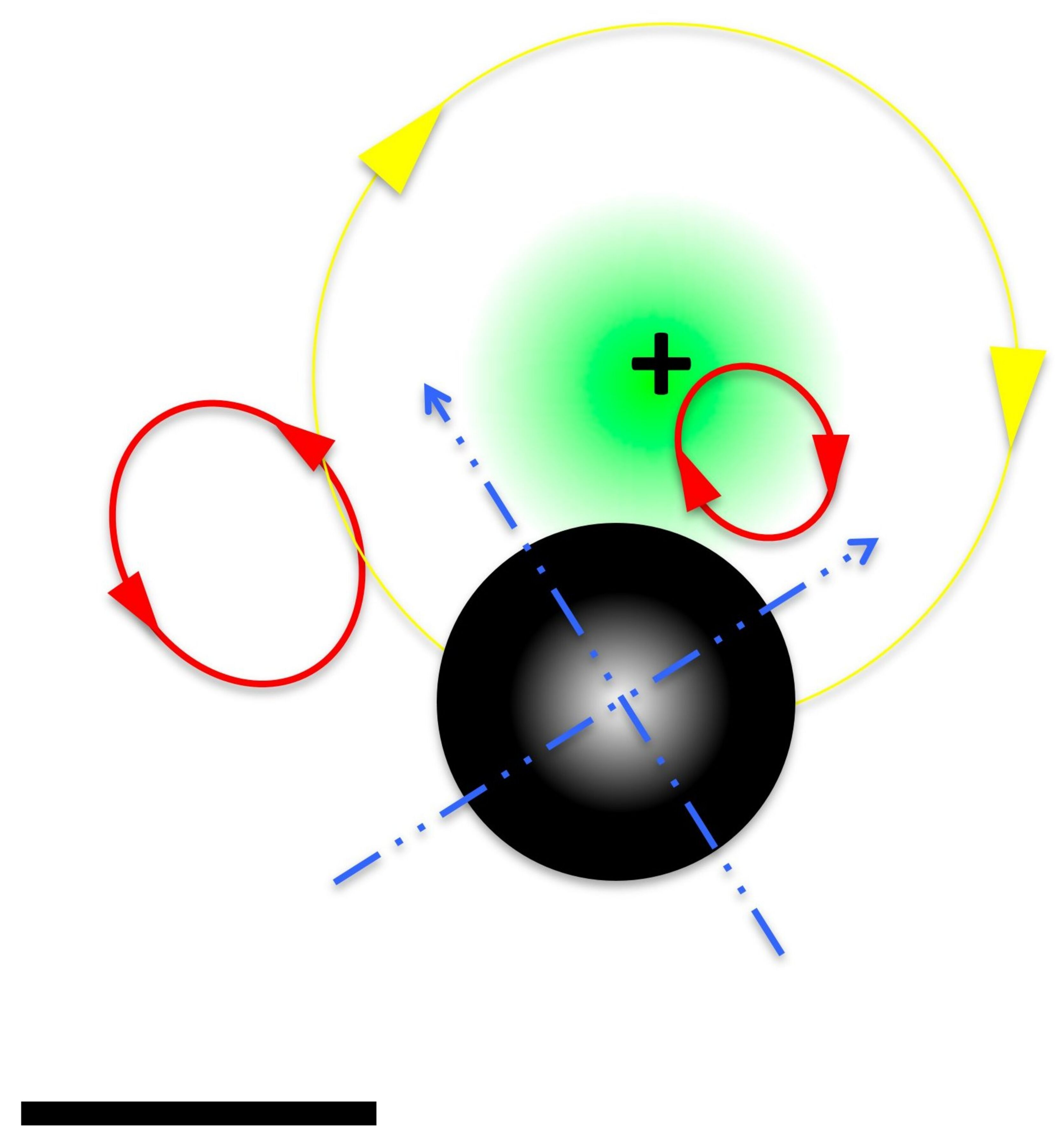}
\caption{Vortex pair forming around an orbiting hot bead. This schematic depicts the situation shown in the Supporting Information Video S6 of Ref.~\cite{Girot2016}. Tracers reveal a first vortex ahead of the bead and a second vortex inside the orbit (red loops). The former rotates anticlockwise whereas the latter rotates clockwise, just like the hot sphere on its track (yellow circle). The black cross marks the center of the vertical laser beam sketched by the green halo. Experimental parameters (Ref.~\cite{Girot2016})\,: beamwaist $\omega_0=6.3~\mu$m, laser power $\mathrm{P}=28$~mW. Scale bar: $5~\mu$m.}
\label{fig:Fig1} 
\end{figure}

The present work is motivated by recent observations of the dynamics of a laser\,--\,heated microsphere at the water\,--\,air (WA) interface~\cite{Girot2016, Zhong2017}. The authors of~\cite{Girot2016} report that, at sufficiently high power, the microsphere orbits the laser beam on quasicircular closed tracks. Most interestingly, they also observe that counter\,--\,rotating vortex pairs escort the hot sphere along its trajectory (Fig.\,\ref{fig:Fig1}). No stable orbits but radial oscillations of the particle arise in~\cite{Zhong2017} under strong heating conditions. This echoes the situation depicted in~\cite{Hauser2018}, except that capillary\,---\,instead of optical forces\,---\,act as restoring forces. Capillarity combines with Marangoni driving to sustain oscillations, much like in Ref.~\cite{Zhong2017}. One major feature common to both studies~\cite{Girot2016, Zhong2017} is that the light\,--\,absorbing microbead behaves as a self\,--\,propelled particle under the action of thermocapillarity.

The present study is the natural continuation of the experiments by Girot \textit{et al.}~\cite{Girot2016}, with the goal to unravel the propulsion mechanism of the hot sphere (Fig.\,\ref{fig:Fig1}). However, characterizing the multipolar flows induced by a hot particle moving at the WA interface is an arduous task. We consider a simpler approach which consists in fixing the heat source at the interface whilst monitoring the convective flows that develop around it. Despite its fundamental interest to interfacial hydrodynamics, this situation has been given only little consideration so far. Among the rare contributors to the problem, Bratukhin and Maurin derived an analytical solution of the coupled Navier\,--\,Stokes and heat transport equations in the nonlinear Marangoni regime~\cite{BratukhinAndMaurin1967}. Later, Shtern and Hussain probed the stability of axisymmetric flows in response to azimuthal perturbations~\cite{ShternAndHussain1993}. On the experimental side, Mizev \textit{et al.}~\cite{Mizev2005} revisited early works pertaining to the structure and stability of surface tension driven flows.\\

The remainder of the paper is structured as follows. We describe the system and its operating principle in Sec.~\ref{DescriptionOfTheSystem}. Next comes the Materials and Methods' Sec.~\ref{MaterialsAndMethods}. We report our experimental observations in Secs~\ref{BaseFlowState} and~\ref{QuadrupolarFlow}. We first consider the base flow state which is simply axisymmetric around the heat source (Sec.~\ref{BaseFlowState}). This type of flow is solely observed under low heating. We will see that experiments yield direct evidence for an elasticity of the interface. What happens when heating is increased is the matter of Sec.~\ref{QuadrupolarFlow}. We show that the radial symmetry of the low\,--\,power pattern breaks down as the flow self\,--\,organizes into counter\,--\,rotating vortex pairs. Salient characteristics of the multipolar flows are then identified. Finally, we summarize and discuss the main outcomes in Sec.~\ref{Discussion}.

\section{Description of the system}
\label{DescriptionOfTheSystem}

\begin{figure}
\centering
\includegraphics[width=0.8\columnwidth]{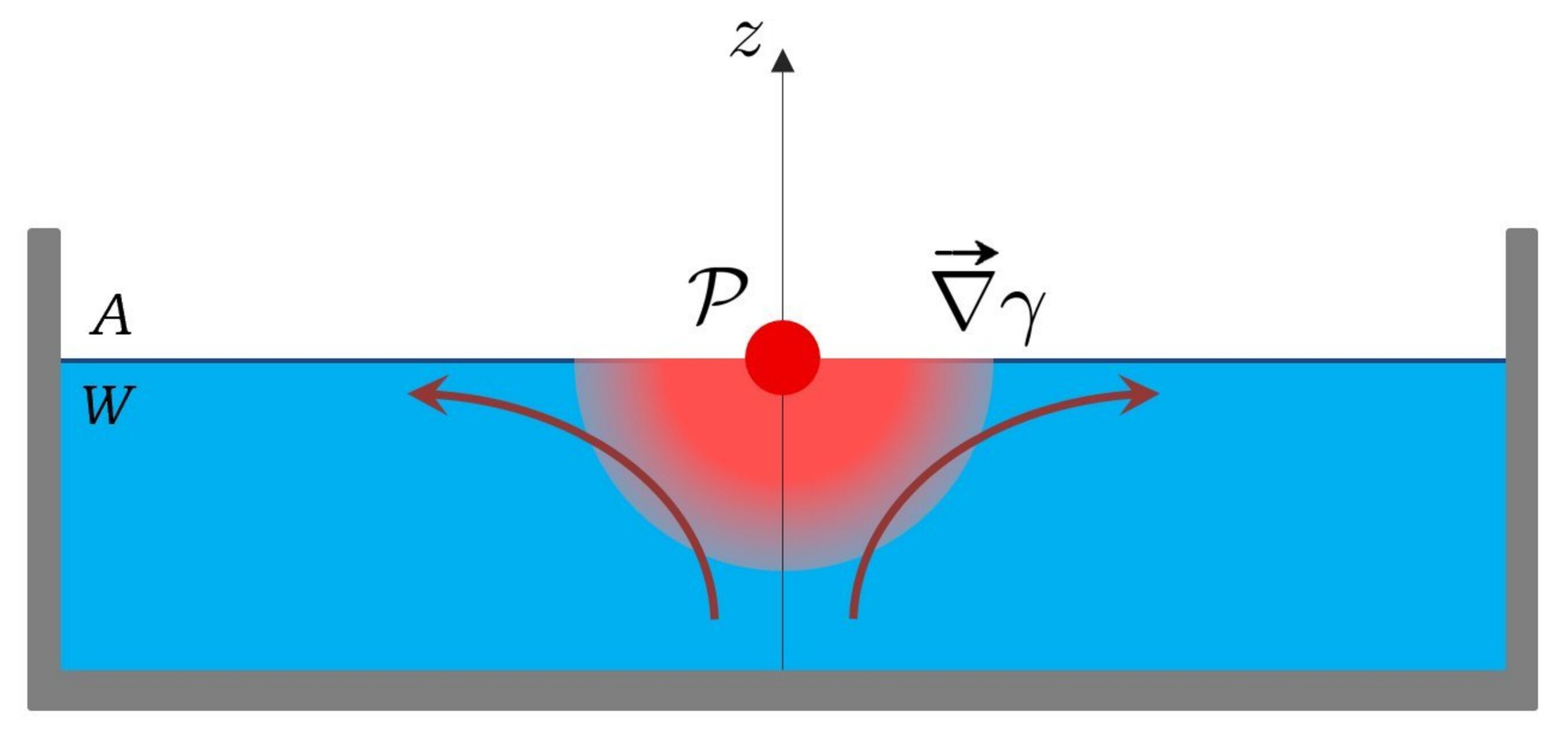}
\caption{Schematic of the system (not to scale). $W$: water; $A$: air; $\mathcal{P}$: laser power; $\protect\vv{\nabla}\gamma$: surface tension gradient. The arrows indicate the direction of the thermocapillary flow.}
\label{fig:Fig2}
\end{figure}

Water fills a cylindrical cell of radius $R$ and height $H$. The $z$\,--\,axis coincides with the axis of revolution of the container. It is oriented upwards, with unit vector $\mathbf{e}_z$.
The principle of our experiment is the following: a hot bead of radius $a$, which is kept fixed as described in Sec.~\ref{MaterialsAndMethods}, sits in partial wetting across the WA interface located at $z=0$ (Fig.\,\ref{fig:Fig2}). Because of thermal gradients in the vicinity of the hot bead, Marangoni flows appear and point towards the colder edges of the vessel. Since we focus on the evolution of the convective flow surrounding the hot bead, the main tunable parameter here is the laser power $\mathcal{P}$. The technical features of our in\,--\,house experimental setup are specified in the following section.\\

\section{Materials and methods}
\label{MaterialsAndMethods}

\subsection{Sample cell}
\label{SampleCell}

The sample cell is a $2R=22.8\,\mathrm{mm}$ wide $\times$ $H=3\,\mathrm{mm}$ high cylindrical quartz cuvette purchased from Thuet France. After a 24h cleaning in sulfochromic acid, the latter is thoroughly rinsed, then filled, with Millipore Milli-Q water. A lid drilled with a $12\,\mathrm{mm}$ circular hole in its center is placed a few mm above the sample cell to limit the pollution of the interface by air contaminants, prevent massive evaporation and avoid surface agitation caused by air turbulence.

\subsection{Hot bead and power supply system}
\label{HotBeadAndPowerSupplySystem}

A glassy carbon bead (Alfa Aesar), whose diameter slightly varies from one experiment to another around $2a=300~\mu$m, is stuck onto the extremity of an optic fiber using an UV curing adhesive (NOA 65). To this end, the fiber has been preliminary stripped up to the cladding. Carbon is prized here for its strong capacity to absorb the incident laser light while resisting photodegradation.

A $532\,\mathrm{nm}$ green laser beam generated by a Quantum Laser Opus source is guided towards the glassy carbon bead by a single\,--\,mode optic fiber (F-SA-C Newport). In our experiments, the laser power $\mathcal{P}$ ranges from a few $\mathrm{mW}$ to a few tens of $\mathrm{mW}$. Because of injection losses within the fiber optic coupler (transmission losses through the fiber are marginal), the power $\mathcal{P}$ impinging on the bead is only a fraction of the power $\mathcal{P}_\mathrm{em}$ emitted by the laser source. In our practical conditions, we have $\mathcal{P}=\epsilon\,\mathcal{P}_\mathrm{em}$ with a coupling efficiency $\epsilon\approx 20\%$. The latter estimate is obtained by measuring with a power meter (Spectra\,--\,Physics SP404) the power of a divergent laser beam which fans out from the fiber at an angle of approximately $30^{\circ}$.

The excess temperature on the surface of the bead $\textDelta T\,\,\dot{=}\,\,T(r=a)-T_0$ ($T_0=25^{\circ}$C, room temperature) reads $\textDelta T=\alpha\left(\mathcal{P}/2\pi\kappa a\right)$ ($\kappa=0.6\,\,\mathrm{W}.\,\mathrm{m}^{-1}.\,\mathrm{K}^{-1}$, water thermal conductivity in standard conditions)~\cite{Wurger2014}. In the latter expression, the prefactor $\alpha<1$ accounts for the fact that only a fraction of the incident power $\mathcal{P}$ converts into heat within the body of the carbon particle. What remains of the optical power is reflected and scattered by the fiber\,--\,bead junction, making the hot sphere radiant (photograph on top of Fig.\,\ref{fig:Fig3}). Since the value of $\alpha$ is not known, we cannot calculate the temperature of the bead using the above expression. However, pushing the system to a maximum power $\mathcal{P}_\mathrm{max}\approx 130\,\mathrm{mW}$ (beyond this value, there is a serious risk of damage) makes small bubbles nucleate on the bead's surface, meaning that we are close to the water boiling temperature $T_b$. Note that in routine conditions the setup is operated at powers far below this limit. Based upon bubbling, and performing a linear interpolation between $T_b$ and $T_0$, we estimate that $\textDelta T$ ranges between a few $\mathrm{K}$, in the axisymmetric flow regime, and $40\,\mathrm{K}$ when the azimuthal instability is fully developed. An accurate measurement of $\textDelta T$ might be obtained by thermography using a dedicated infrared microscopy hardware, however, the latter has not been set up yet.

An alternate way to estimate the temperature of the carbon sphere is to perform a numerical simulation of the fiber\,--\,bead geometry with the bead in partial wetting configuration at the water\,--\,air interface. The simulation shall compute the resulting flow field, which first requires solving the full hydrodynamic problem (including heat transport from the surface of the carbon sphere). This objective, beyond the scope of the present article, will be part of a future publication~\cite{LoudetInProgress}.

\subsection{Bead/water contact}
\label{BeadWaterContact}

\begin{figure}
\centering
\includegraphics[width=0.85\columnwidth]{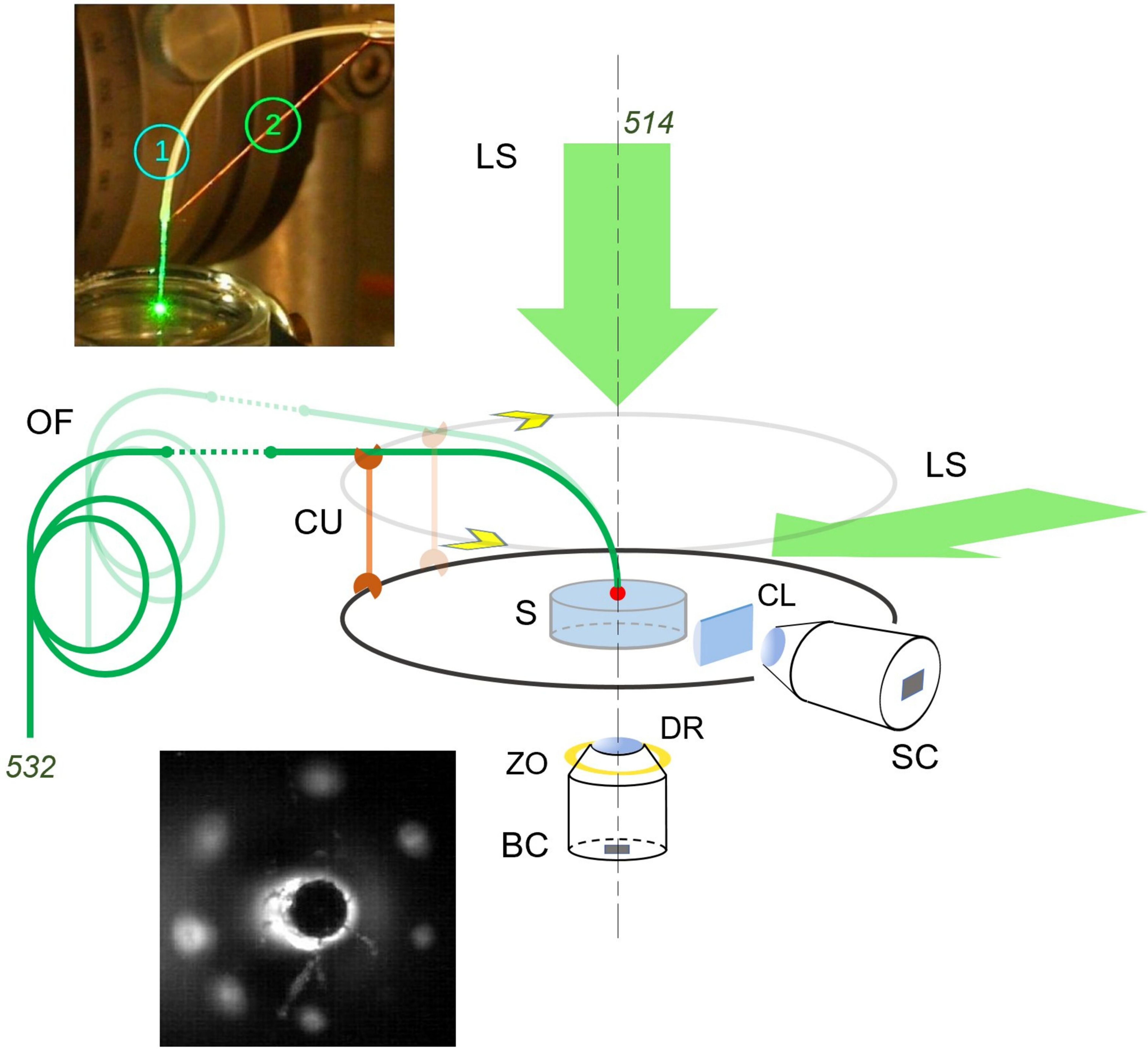}
\caption{Experimental setup (not to scale). S, sample; OF, optic fiber guiding the $532\,\mathrm{nm}$ green laser beam that heats the bead; CU, coupling unit to rotate the set \{optic fiber\,$+$\,bead\} on the circular guide rail, as shown by the transparent part of the schematic; LS, $514\,\mathrm{nm}$ green laser sheets; \{BC\,$+$\,ZO\}, bottom camera\,$+$\,zoom objective; DR, diode ring; \{SC\,$+$\,CL\}, side camera\,$+$\,correction lens. The top insert is a photograph of the optic fiber (1) bent by a thread of adjustable tension (2) that ensures the axis of the bead is perpendicular to the WA interface. The bottom insert shows the ring of light\,--\,emitting diodes as imaged on the video screen. The carbon sphere is the dark disk well visible in the middle.}
\label{fig:Fig3} 
\end{figure}

In order to restrict the number of factors influencing the system, we endeavor to keep the interface as flat as possible. To this end, the cuvette is prefilled to the brim and water is then progressively removed. In such a way, a flat interface is obtained far from the heat source by contact line pinning on the sharp edges of the cuvette. For the axis of the fiber to be oriented perpendicularly to the surface while passing through the protective lid, the fiber is bent using a thread of adjustable tension tied around its coating and stretched between the fixation point and the fiber mounting plate (Fig.\,\ref{fig:Fig3}, top insert).

The bead is displaced with a $xyz$ translation stage till being partially immersed in water. The tension exerted by the fiber on the bead leads to the formation of a steep meniscus which, for the same reason as above, shall be minimized as much as possible. Simply stated, we want the interface to remain flat around the carbon sphere. The deformation of the interface is monitored using a ring encrusted with light\,--\,emitting diodes that we position beneath the cuvette. The light from the diodes reflects off the water surface yielding a luminous circle on the video screen (Fig.\,\ref{fig:Fig3}, bottom insert). We then finely tune the altitude of the \{fiber\,$+$\,bead\} unit in such a way that the circle shrinks down and is no longer visible on the image. The latter tuning is accurate within a few $\mathrm{µm}$. Because of water evaporation around the hot bead, the procedure must be repeated several times within a single experimental run which usually lasts for one hour.

\subsection{Flow visualization}
\label{FlowVisualization}

Flow visualization is based on laser sheet tomography. This technique requires seeding the liquid with tracers. Polystyrene beads from Magsphere (density $\rho_\mathrm{PS}=1.05$) with a diameter of $5.1\,\mathrm{\mu m}$ serve here as tracer particles. We checked with a dedicated experiment~\cite{NoteOnTracers} that the latter are neutral towards the investigated flows, a basic requirement any suitable tracer shall meet. Note that a volume fraction of these beads as minute as $\varphi_v\sim 10^{-5}$ is sufficient to ensure proper flow visualization. In the plane of a laser sheet, tracers emit a fluorescent light that is sensed by video cameras. In the end, streamlines appear on time\,--\,lapse pictures as streaks of light left by tracers along their trajectories.

A couple of laser sheets powered by a $514\,\mathrm{nm}$ green laser source (Genesis CX 514\,--\,2000 STM from Coherent) allows for cutting the sample cell along selected planes\,: a horizontal laser sheet is positioned at a short distance beneath the interface ($z\approx -0.1\,\mathrm{mm}$) to visualize the surface flows while a vertical laser sheet, tangent to the bead, yields almost diametral cuts of the bulk flows~\cite{NoteOnLaserSheetsPositions}.

A key element to capture properly the structure of the flows is to rotate the side camera around the cell so as to align the axis of the objective with a symmetry axis of the vortex pattern. We choose the alternate option whereby the camera is fixed and the vortex pattern can be rotated. In practice, the set \{fiber + bead\} is mounted on a circular guide rail (Fig.\,\ref{fig:Fig3}).

Flows are recorded from the bottom and the side with CCD video cameras (Fig.\,\ref{fig:Fig3}). Depending upon the needs, we use either an EO\,--\,1312M or a pco.pixelfly camera. The resolution of the video pictures is limited by the thickness of the laser sheets ($e\approx 50~\mu$m), the defaults of the optical imaging hardware (essentially astigmatism) and the size of the camera pixels. A spatial resolution of about $10~\mu$m is estimated empirically from the video images. The latter value holds in the region $\{r<1\,\mathrm{cm}\}$ not too far from the vertical $z$\,--\,axis where details are not significantly blurred by astigmatism (a correction lens is placed between the sample and the side camera to fix this issue as much as possible, see Fig.\,\ref{fig:Fig3}). Note that images are exploited only within the latter viewing area. Tracers located at larger distances appear motionless. The time resolution is set by the frame rates of the cameras which, depending upon the case, are operated from 2 up to 30 frames per second.

\subsection{Flow velocity measurements}
\label{FlowVelocityMeasurements}

Flow velocities are measured either by single particle tracking or particle image velocimetry (PIV)~\cite{Raffel2007}\,--\,\cite{ThielickeAndStamhuis2014}. In this study we use PIV only within vertical cuts. The method can be applied to horizontal cuts, but not close to the WA interface due to large variations of longitudinal velocities over the thickness of the laser sheet~\cite{NoteOnPIVInaccuracyOnHorizontalCuts}. Below we uncover correlations between the morphology of the surface flows and the interfacial dynamics by comparing surface with subsurface velocities. Such an analysis relies on the ability to distinguish between surface and bulk tracers while tracking them on vertical cuts\,: we exploit the fact that particles at a small distance below the free surface ($z<0$) display well discernible mirror images resulting from their reflection off the interface, whereas particles found at the surface ($z=0$) appear as single bright spots. This property is very helpful for locating the interface and hence measuring with pixel accuracy ($\delta z\approx 10~\mu$m) the depths at which bulk tracers travel.

\section{Base flow state}
\label{BaseFlowState}

\begin{figure}
\centering
\includegraphics[width=0.8\columnwidth]{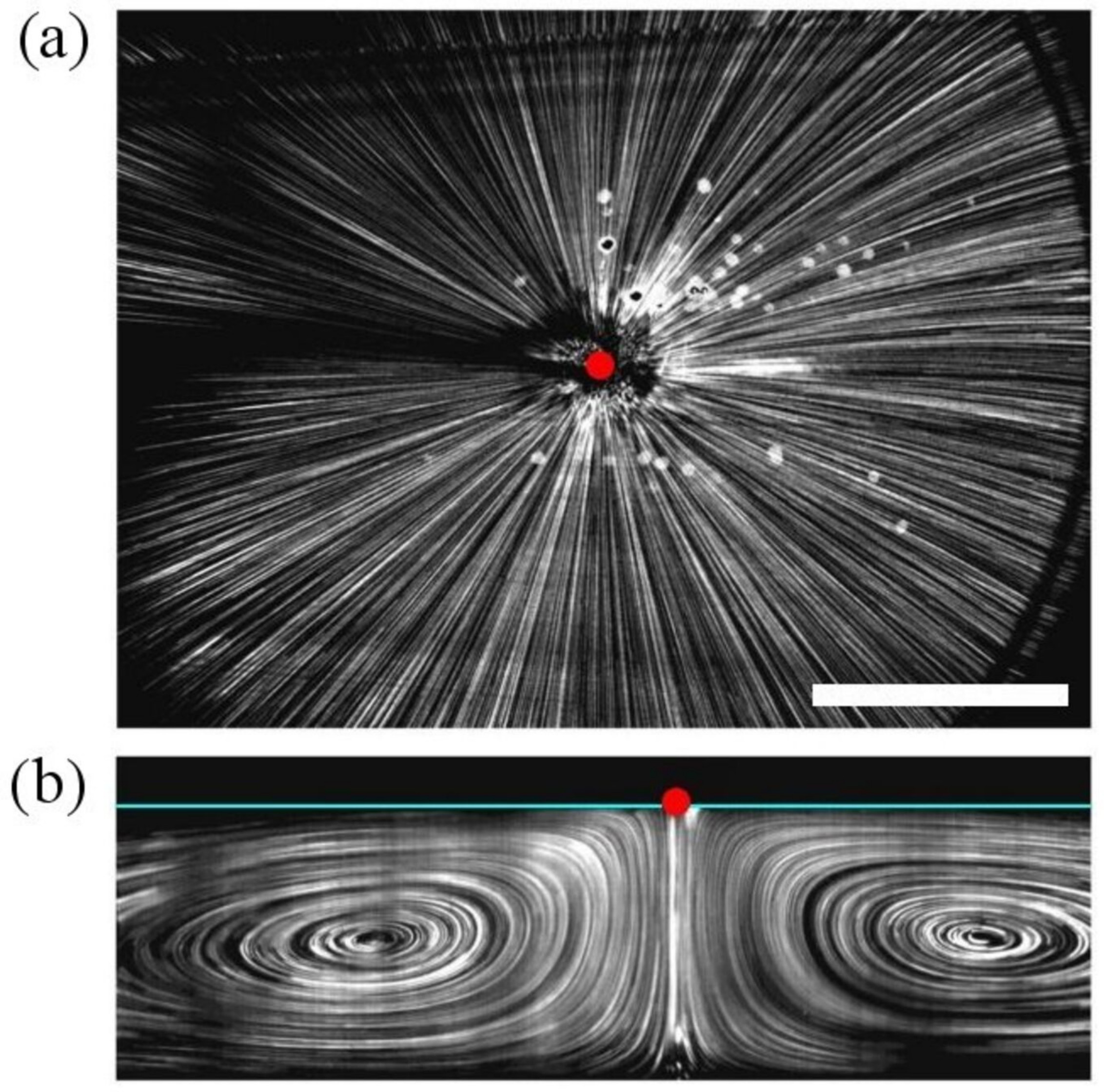}
\caption{Base torus. (a) Centrifugal surface flow. (b) Flow in a vertical cross\,--\,section. The cyan line marks the position of the interface. The hot bead ($2a\approx 300\,\mathrm{\mu m}$) is sketched by a red disk on both figures. The bright spots visible on Fig.\,4(a) around the hot bead are due to clusters of tracers settled on the cell floor. Scale bar (common to both figures)\,: $3\,\mathrm{mm}$.}
\label{fig:Fig4} 
\end{figure}

The thermocapillary flow observed under low heating, typically for $\mathcal{P}\sim 1\,\mathrm{mW}$, has the topology of a torus whose revolution axis coincides with the axis of the source (Fig.\,\ref{fig:Fig4})~\cite{NoteOnImperfectAxisymmetry}. This steady flow is the base flow state.

\subsection{A nearly solid interface}
\label{ANearlySolidInterface}

\begin{figure}
\centering
\includegraphics[width=0.8\columnwidth]{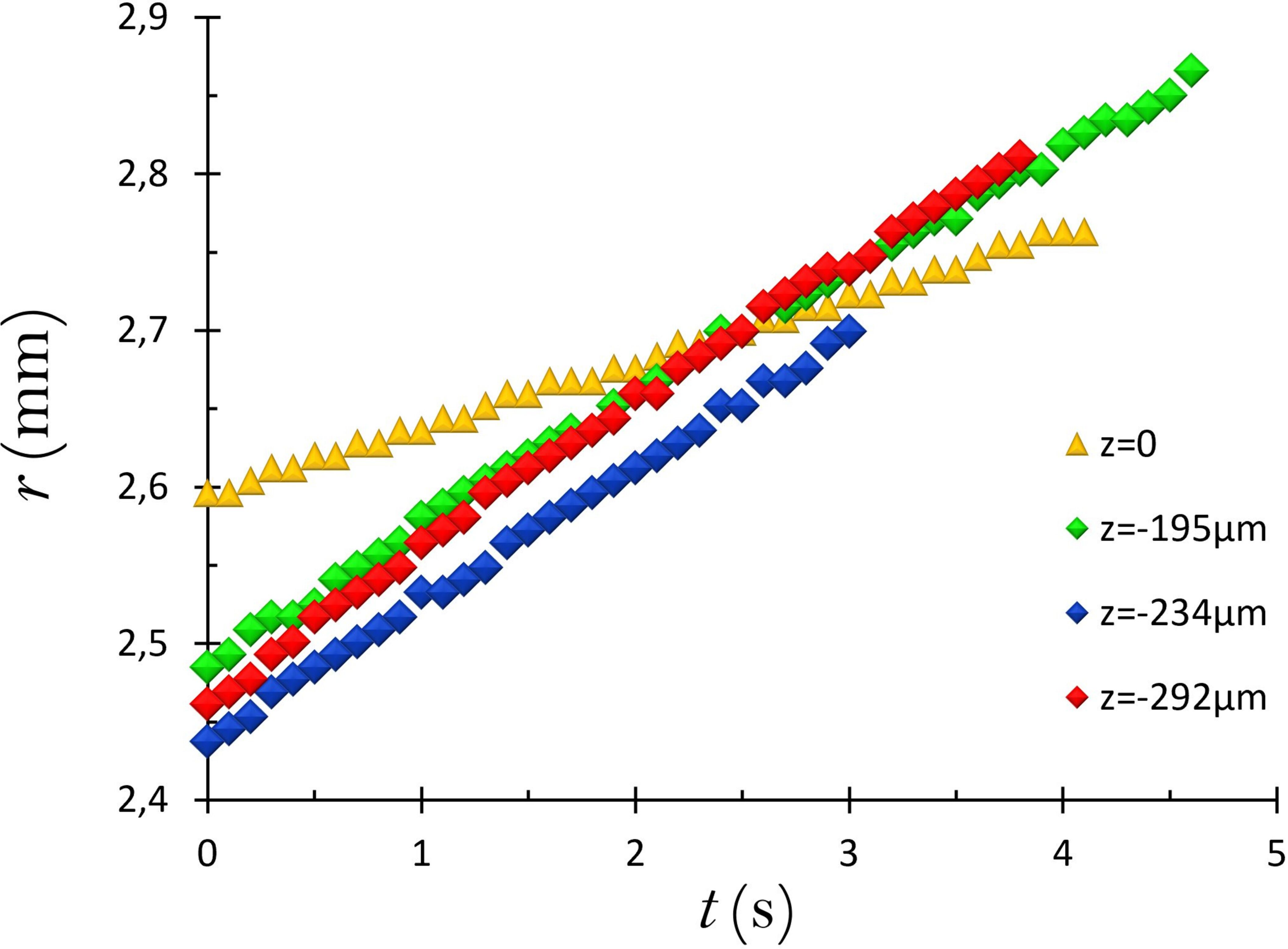}
\caption{Time evolution of the radial distance $r$ to the hot bead of surface and subsurface tracers in a cross\,--\,section of the base torus. The travelling depth $z$ of each tracer is given next to its symbol. Position uncertainty: $\delta z=\pm 10\,\mathrm{µm}$. The surface tracer ($z=0$) has a velocity $v=42~\mu$m/s, while subsurface tracers located at $z=-195,-234,-292~\mu$m move at $v=81,86,92~\mu$m/s, respectively. Velocity uncertainty: $\delta v=\pm\,5~\mu$m/s. $\mathcal{P}=20\,\mathrm{mW}$.}
\label{fig:Fig5} 
\end{figure}

We compare the velocity of interfacial tracers with that of tracers situated in a shallow layer which extends down to a few tenths of a mm underwater. Measurements are made through direct particle tracking in a cross\,--\,section of the base torus. In practice, the centrifugal motion of four tracers located about $2.5\,\mathrm{mm}$ far from the hot bead is tracked over time. One tracer sits at the interface ($z=0$) while the other three lie at shallow depths ($z<0$). The corresponding trajectories are plotted on Fig.\,\ref{fig:Fig5}. Direct comparison of the slopes clearly reveals that subsurface flow velocities are higher than surface ones. The latter hierarchy among velocities is the imprint of a `nearly solid interface' imposing a `quasi no\,--\,slip' boundary condition for the fluid velocity. We will clarify the meaning of the expression `quasi no\,--\,slip' in the discussion (Sec.\,\ref{Discussion}).

\begin{figure}
\centering
\includegraphics[width=0.8\columnwidth]{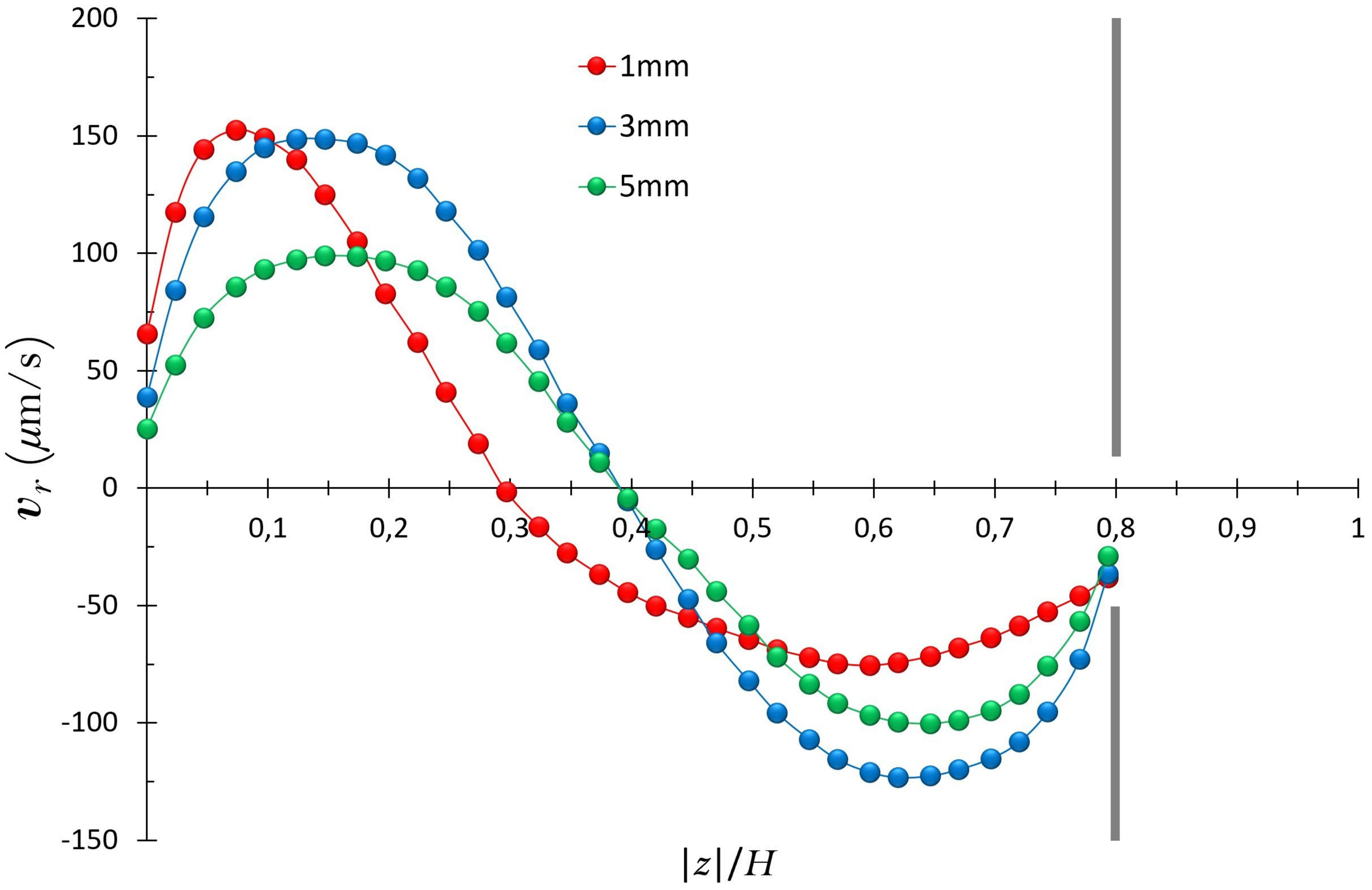}
\caption{Evolution of the radial velocity $v_r$ as a function of depth $z$ (dimensionless unit, $H=3\,\mathrm{mm}$\,: height of the cell). Different colors correspond to different distances to the heat source. The solid lines are a guide to the eye. The area to the right of the thick grey line, for which $0.8\leq|z|/H\leq 1$, corresponds to a region of the flow that is out of the camera's field of view. Indeed, the correction lens (Fig.~\ref{fig:Fig3}) is arranged in such a way that the aiming line of the side camera is slightly tilted up. The latter configuration ensures that the interface is correctly imaged but, at the same time, it prevents us from visualizing the flow near the cell floor. $\mathcal{P}=8\,\mathrm{mW}$.}
\label{fig:Fig6} 
\end{figure}

Below, we show data obtained by PIV in a vertical cross\,--\,section of the base torus. Fig.\,\ref{fig:Fig6} displays the radial velocity as a function of depth. As already evidenced by particle tracking above, the flow velocity is highest at finite depth and not at the interface as may have been expected\,: the maximum radial velocity $|v_{r,\,\mathrm{max}}|$, comprised between $75~\mu$m/s and $150~\mu$m/s, is reached at a depth $z_\mathrm{\,max}$ such that $0.05<|z_\mathrm{\,max}|/H<0.2$ in the explored range of radial positions. Interestingly, we remark that $z_\mathrm{\,max}$ increases with increasing distance to the heat source. Furthermore, a recirculation flow arises in the bulk owing to mass conservation, as revealed by the flow velocity systematically changing sign below some critical depth $z_\mathrm{\,inv}$ (here $0.3<|z_\mathrm{\,inv}|/H<0.4$). The fact that subsurface flow velocities are higher than surface ones with the bulk velocity going through a maximum at a finite depth (plus mass conservation) explains the universality of the obtained S\,--\,shaped profiles which are found in many studies on Marangoni convection in finite size systems (e.g., Figs 3 in~\cite{Favre1997,Shmyrov2019}).

\subsection{Centrifugal motion of a surface tracer}
\label{CentrifugalMotionOfASurfaceTracer}

\begin{figure}
\centering
\includegraphics[width=0.8\columnwidth]{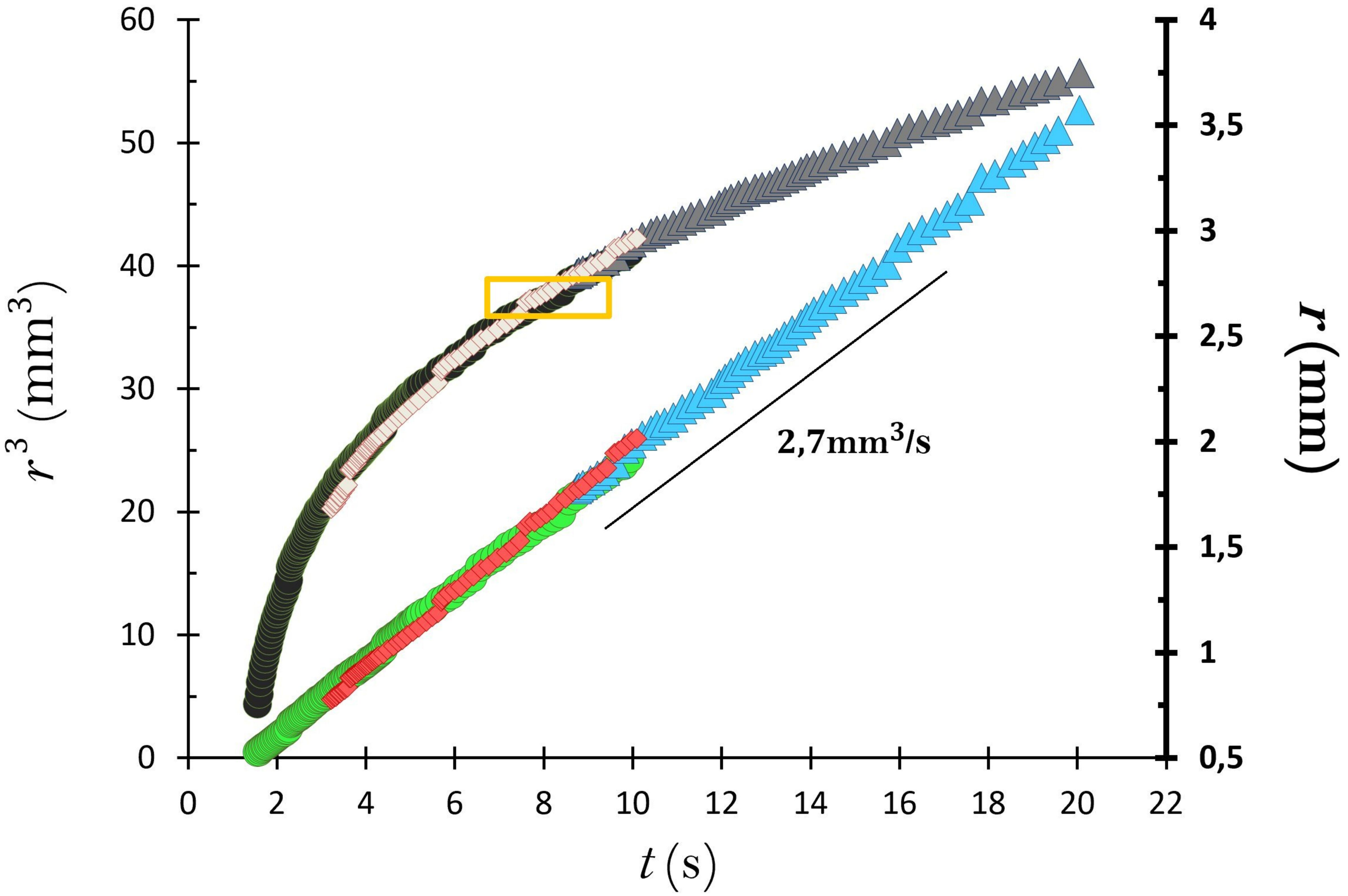}
\caption{Cube of the radial distance $r$ of surface tracers to the hot bead as a function of time $t$ (brightly colored line). Each colored section represents the trajectory of a tracer. A secondary ordinate axis is added to plot the corresponding $r=f(t)$ curve (grey tones). The orange box marks the data range of the surface tracer displayed in Fig.\,\ref{fig:Fig5}. $\mathcal{P}=14\,\mathrm{mW}$.}
\label{fig:Fig7} 
\end{figure}

We now aim at characterizing the centrifugal motion of surface tracers. The trajectories of tracers located on the interface at varying distances from the heat source are `bound together' so as to reconstruct a full radial trajectory~\cite{NoteOnTrajectoryReconstruction}. This amounts to time\,--\,shifting tracers' positions until a single representative trajectory is generated out of the trajectories of individual particles. Note that this operation relies on the assumption of flow steadiness. Fig.\,\ref{fig:Fig7} clearly reveals a law of motion $r\sim t^{1/3}$. As will be discussed in Sec.\,\ref{Discussion}, this scaling departs from the behavior expected for a pure thermocapillary flow.

There seems to be a contradiction between the $r\sim t^{1/3}$ behavior of surface tracers evidenced in Fig.\,\ref{fig:Fig7} and the apparently linear behavior suggested by Fig.\,\ref{fig:Fig5}. However, the range $2.59\,\mathrm{mm}\leq r\leq 2.77\,\mathrm{mm}$ of radial distances spanned by the surface tracer of Fig.\,\ref{fig:Fig5} represents only a little portion of the $r=f(t)$ curve plotted in Fig.\,\ref{fig:Fig7}. The corresponding area is delineated by a rectangle in Fig.\,\ref{fig:Fig7}. One can see that the latter is  located in a region where $r$ is linear in $t$ to a first approximation.

\subsection{Response to laser shutdowns}
\label{ResponseToLaserShutdowns}

In the following, we probe the dynamic response of the base torus to laser shutdowns. In practice, the laser is suddenly switched off using a beam stop. Note that the time $\textDelta t_\mathrm{on\rightarrow off}$ ($\textDelta t_\mathrm{off\rightarrow on}$) during which the laser is on (resp.\,\,off) must be chosen long enough for the flow to reach a steady state (resp.\,\,vanish). We set $\textDelta t>20\,\mathrm{s}$ as determined from preliminary tests.

We note the onset of a short\,--\,lived ($\approx 10\,\mathrm{s}$) centripetal motion of surface tracers at the very moment $t_\mathrm{off}=0$ when the laser is turned off (see Supporting Information Video V1). The latter retraction phenomenon strongly suggests that the WA interface behaves as an elastic membrane because of the probable presence of adsorbed surface\,--\,active species. We actually have evidence, from ongoing numerical simulations which will be the matter of a forthcoming publication~\cite{LoudetInProgress}, that the aforementioned reversed motion of surface tracers after laser shutdown is a very likely indication of the presence of adsorbed surfactants at the water\,--\,air interface.

\begin{figure}
\centering
\includegraphics[width=0.8\columnwidth]
{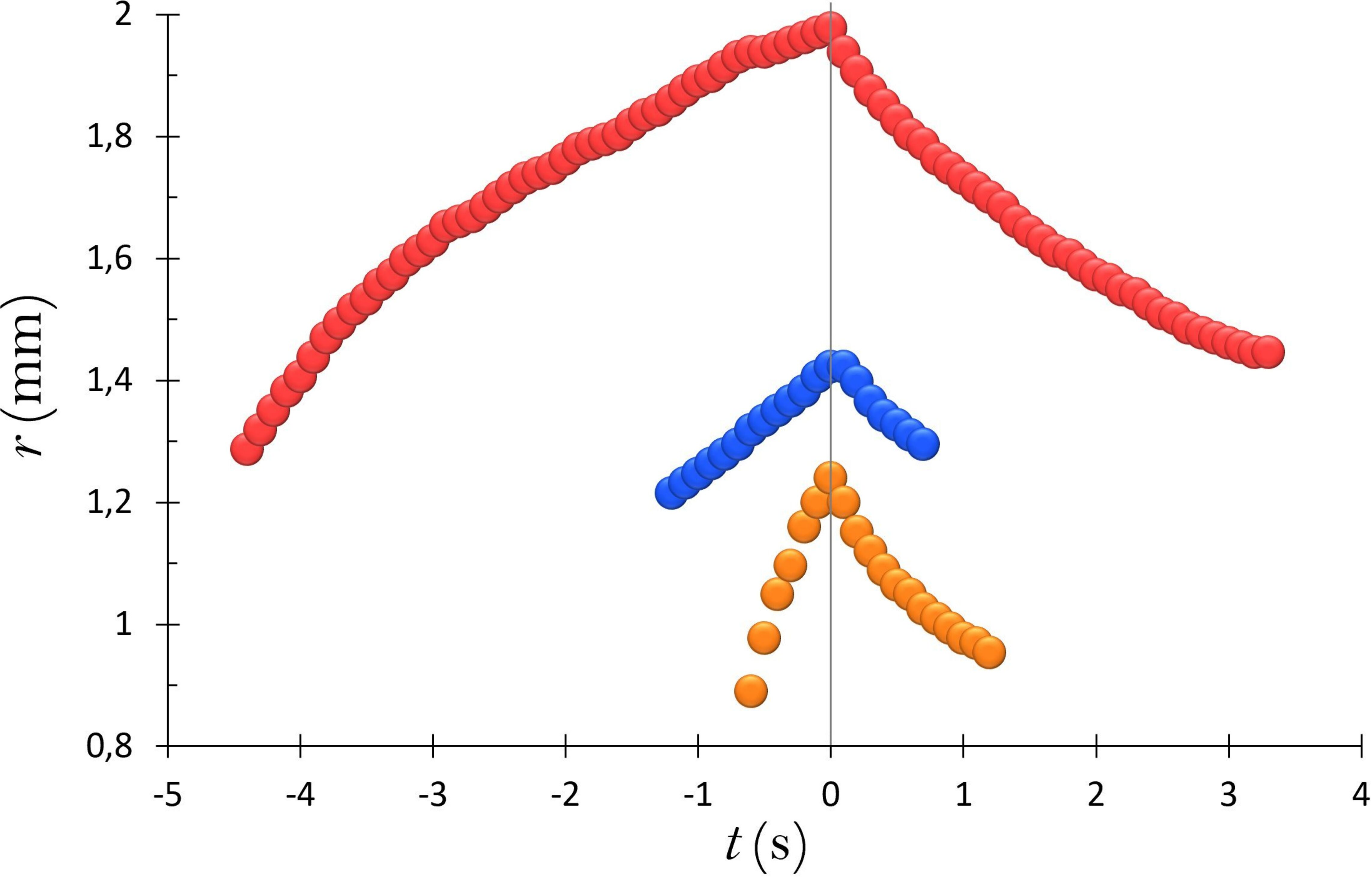}
\caption{Evidence of interface elastic retraction. The vertical line marks the instant $t_\mathrm{off}=0$ of the laser shutdown. Different colors correspond to different tracers. $\mathcal{P}=20\,\mathrm{mW}$.}
\label{fig:Fig8} 
\end{figure}

Elastic retraction is definitely confirmed by tracking surface tracers along their paths. Fig.\,\ref{fig:Fig8} shows how the radial velocity of interfacial tracers sharply (angular point) reverses at $t_\mathrm{off}=0$\,: just after the laser extinction, tracers start moving in the opposite direction as revealed by a change in the sign of the slope measuring the flow velocity. We check that the closer the particle to the heat source, the sharper the peak on the curve, that is the higher its pre\,--\,and post\,--\,shutdown speed.

\section{Quadrupolar flow}
\label{QuadrupolarFlow}

In a typical experiment we progressively increase the laser power $\mathcal{P}$, starting with the axisymmetric base flow presented in Sec.~\ref{BaseFlowState}. Beyond a critical power $\mathcal{P}^\ast$, the flow looses its radial symmetry and self\,--\,organizes into vortex pairs. We observed one vortex pair (a dipole) or two pairs (a quadrupole), with no obvious trend that a dipole shows up before a quadrupole when the power is slowly increased. In most cases, the flow was observed to transition directly from axisymmetric to quadrupolar. One strong practical limitation comes from the fact that the properties of the WA interface evolve significantly within about one hour. As will be discussed in Sec.~\ref{Discussion}, the problem is due to contamination by surface\,--\,active species. Even though interface pollution involves minute amounts of contaminants it cannot be avoided, at least under the experimental conditions set up in the present work, whatever the care taken in preparing the samples. Interface contamination increases within minutes and its amplitude definitely varies between distinct experimental runs. Consequently, only a rough estimate of $\mathcal{P}^\ast$ can be made within a single run, and the latter may strongly vary between different runs. Typical values of $\mathcal{P}^\ast$ range between $20\,\mathrm{mW}$ and $30\,\mathrm{mW}$.

The transition from the axisymmetric regime to the multipolar flow is qualitatively reversible, which means that reducing the power below some value $\mathcal{P}^{\ast\ast}$ brings the system back to the base flow state. However, the value of the crossover power $\mathcal{P}^{\ast\ast}$ is in general lower than $\mathcal{P}^\ast$. Because of the aforementioned issue, it is not possible to tell whether the difference between $\mathcal{P}^\ast$ and $\mathcal{P}^{\ast\ast}$ reflects a truly hysteretic behavior or is simply due to growing contamination (`surface ageing').

\begin{figure}
\centering
\includegraphics[width=0.8\columnwidth]{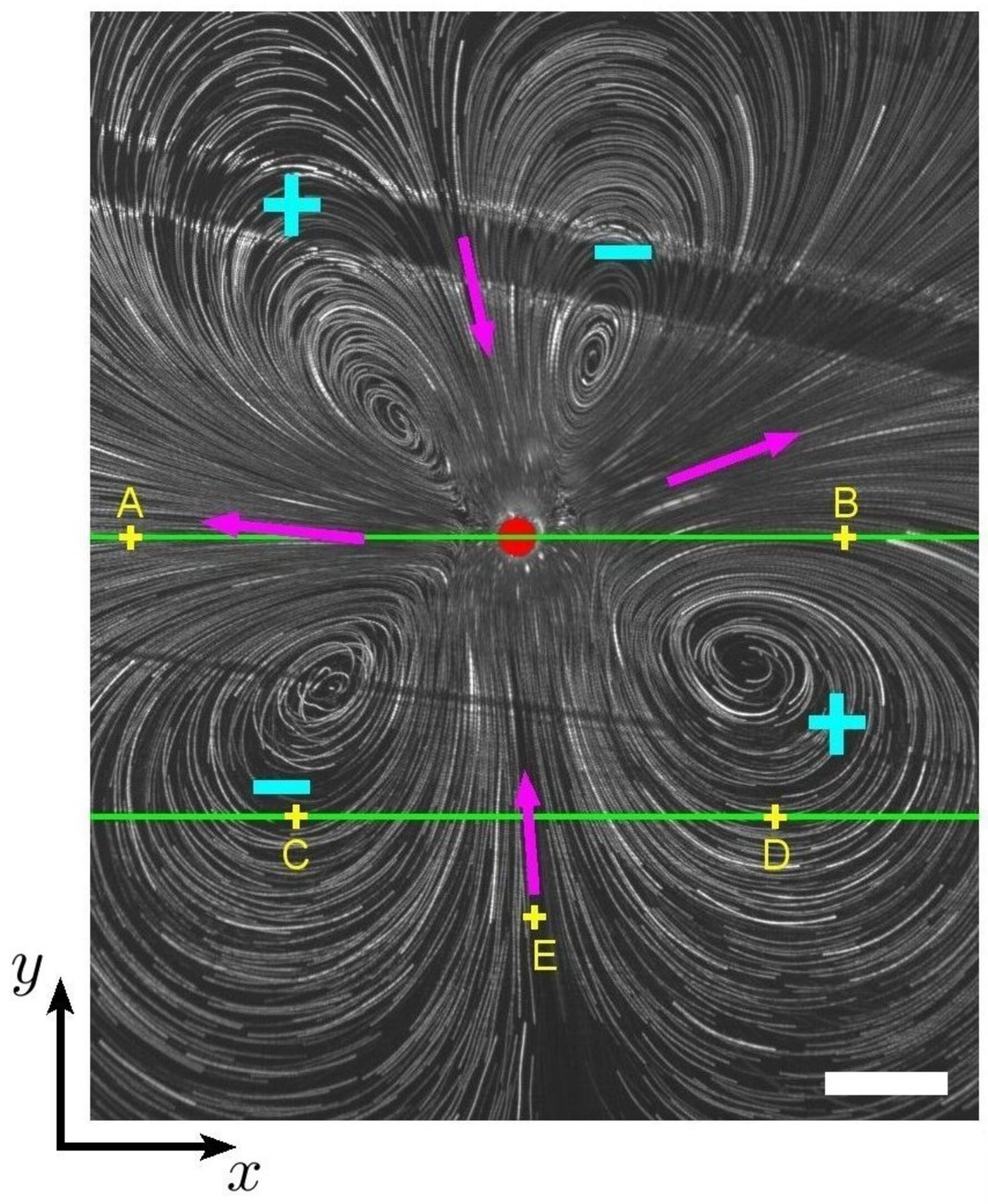}
\caption{Measuring local velocities in various regions of a quadrupolar flow (\textit{top view}). The arrows show the direction of the flow. $+/-$\,: clockwise/anticlockwise vortex rotation. The vertical laser sheets are depicted by green lines. The hot bead ($2a=295\,\mathrm{\mu m}$) is sketched by a red disk. $\mathcal{P}=70\,\mathrm{mW}$. Scale bar\,: $1\mathrm{mm}$.}
\label{fig:Fig9} 
\end{figure}

In general, vortex pairs forming quadrupoles were not equal in size. We noticed that the hierarchy between vortex sizes could nevertheless evolve in time and even reverse. Below we focus on an experiment where a neat quadrupole is observed and is stable enough to perform a thorough analysis of its features. The investigated flow is shown in Fig.\,\ref{fig:Fig9}\,: two counter\,--\,rotating vortex pairs are clearly visible with the four vortices separated by two `channels', one centrifugal and the other centripetal, intersecting at (nearly) right angles.\\

The flow is scrutinized in a couple of cross\,--\,sectional planes\,: a first viewing plane is positioned close to the centrifugal channel while a second plane, parallel to the first one, cuts across the eddies. More precisely, local flow velocities are measured in the immediate vicinity of points A to D. This study is complemented with an estimate of the velocity near point E which lies within the centripetal channel (Fig.\,\ref{fig:Fig9}).

\subsection{A passing interface along the channels}
\label{APassingInterfaceAlongTheChannels}

\begin{figure}
\centering
\includegraphics[width=0.8\columnwidth]{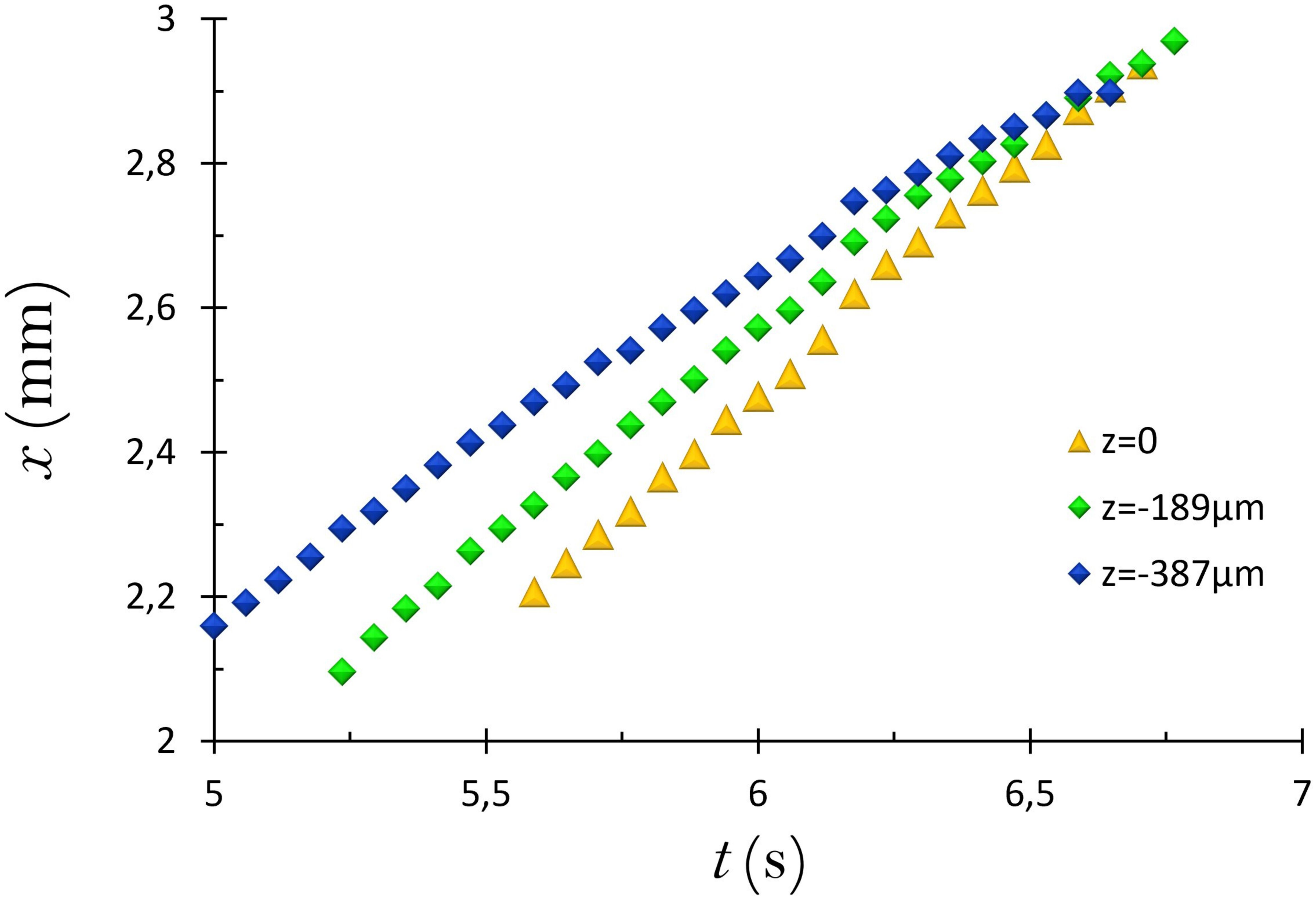}
\caption{Time evolution of the distance $x$ to the hot bead of surface and subsurface tracers in the cut plane AB (Fig.\,\ref{fig:Fig9}). The travelling depth $z$ of each tracer is indicated next to its corresponding symbol. Position uncertainty: $\delta z=\pm 10~\mu$m. The surface tracer ($z=0$) has a velocity $v=666~\mu$m/s, while subsurface tracers located at $z=-189,-387~\mu$m move at $v=601,485~\mu$m/s, respectively. Velocity uncertainty $\delta v=\pm\,40~\mu$m. $\mathcal{P}=70\,\mathrm{mW}$.}
\label{fig:Fig10}
\end{figure}

Tracking surface and subsurface tracer particles in the cross\,--\,section AB reveals that the flow velocity is higher at the interface than in the shallow depth region (Fig.\,\ref{fig:Fig10}). Recall that the opposite occurs in the ground flow state (Sec.~\ref{ANearlySolidInterface}). The interface is here `passing' in the sense that a quasi stress\,--\,free boundary condition sets in along the centrifugal channel.

In the present experiment, the vertical laser sheet is not oriented parallel to the centripetal channel. It is still possible to estimate local velocities along the section of the centripetal channel close to point E (Fig.\,\ref{fig:Fig9}) based on a cut plane tangent to the interface. By doing this, we measure flow velocities about half those reported for the centrifugal channel. We therefore assume a qualitatively similar situation for the centrifugal and the centripetal channels, namely a `passing' interface in both cases.\\

\subsection{Swirling flow region}
\label{SwirlingFlowRegion}

Examining the flow in the subsurface region of the cut plane CD (Fig.\,\ref{fig:Fig9}), we find velocities that are markedly lower than those measured in either the centrifugal or the centripetal channel (Table~\ref{fig:Table1}). Most importantly, in the swirling flow region along the cross\,--\,section CD, subsurface velocities are larger than interfacial ones, contrary to what occurs along the centrifugal channel where an intense surface flow is observed. The interface is here `nearly solid' like in the toroidal state (Sec.~\ref{ANearlySolidInterface}).\\

\begin{table}
\setlength{\extrarowheight}{3pt}
\begin{center}
\begin{tabular}{|C{1.75cm}||C{2.25cm}|}
\hline
$z$ ($\mu$m) & $v$ ($\mu$m/s)\\
\hhline{|=::=|}
0 & 6.6\\
\hhline{|-||-|}
$-75$ & 31 \\
\hhline{|-||-|}
$-94$ & 45\\
\hhline{|-||-|}
$-206$ & 73\\
\hhline{|-||-|}
$-280$ & 122\\
\hline
\end{tabular}
\end{center}
\caption{Centripetal velocity of surface and subsurface tracers as a function of depth $z$ in the cross\,--\,section CD (Fig.\,\ref{fig:Fig9}). Measurements (uncertainty $\delta v=\pm\,3~\mu$m/s) are made in the left half of the laser sheet where the local velocity is positive.}
\label{fig:Table1}
\end{table}
 
To sum up, tracers move along the centrifugal channel at very high velocities up to $v\approx 1\,\mathrm{mm/s}$. The relative increase in the velocity is more pronounced at the surface ($\times 30$) than in the bulk ($\times 14$), and much higher than the laser powers ratio ($\times 3$), while comparing the results of the quadrupolar flow with those of the base torus. The intervortex regions of the quadrupole are subject to strong flows characterized by larger velocities at the surface than in the bulk, in line with a `passing' interface. By contrast, in the vicinity of vortex centers the situation is reminiscent of the `nearly solid' interface reported for the base torus (Sec.~\ref{ANearlySolidInterface}).

\section{Discussion}
\label{Discussion}

The present study yields cogent evidence of the high sensitivity to azimuthal perturbations of the divergent flow driven by a heat source at the water\,--\,air interface. We show how, from a base torus under low heating, the axisymmetry of the flow breaks down into multipoles at sufficiently high powers~\cite{Mizev2005}. The instability takes the shape of a corolla of counter\,--\,rotating vortex pairs that self\,--\,organize periodically all around the source of heat. In this paper, both the toroidal base flow (Sec.~\ref{BaseFlowState}) and the quadrupolar mode of the instability (Sec.~\ref{QuadrupolarFlow}) were investigated.

Most importantly, we observed the elastic response of the interface to laser shutdowns (Sec.~\ref{ResponseToLaserShutdowns}) as evidenced by the reversed motion of surface tracers, from centrifugal to centripetal (Fig.\,\ref{fig:Fig8}). We assume that the elasticity of the water\,--\,air interface stems from the presence of an adsorbed surfactant layer. This is a natural hypothesis since water, due to its high surface tension relative to many common liquids ($\gamma_\mathrm{water}=72.8\,\mathrm{mN}.\,\mathrm{m}^{-1}$ at $20^{\circ}$C), acts as a receptacle for most surface\,--\,active impurities unavoidably present in the environment. 

Water contamination is a long\,--\,standing problem in interface science~\cite{Kim2017, Uematsu2019}. For instance, the pivotal role of surfactants in retarding the motion of rising bubbles has been recognized in~\cite{YbertAndDiMeglio1998}\,--\,\cite{Takagi2011}. Surface contamination is also suspected to affect the shape of `coffee rings' left by evaporating droplets~\cite{Deegan1997}\,--\,\cite{Kim2016}. Interfacial effects become increasingly dominant while downsizing the system owing to magnified surface\,--\,to\,--\,volume ratios. As an example, microfluidic experiments highlighted that a tiny amount of surfactants can severely undermine the drag reduction potential of superhydrophobic surfaces~\cite{FrancoisPeaudecerf2017}. Impurities also have the ability to alter the viscoelastic response of a water\,--\,air interface, as brought to light by AFM measurements~\cite{Manor2008, Maali2017}. Other experiments suggest that surface\,--\,active contaminants can promote the rupture of $\mathrm{µm}$-thick free liquid films~\cite{Neel2018}. Strikingly, the influence of surfactants manifests itself even at the nanoscale\,: the stability of interfacial nanobubbles has been attributed to impurities~\cite{Ducker2009,Das2010}, while nanomolar concentrations of charged contaminants have been invoked to explain anomalous surface tension variations (Jones\,--\,Ray effect) reported for electrolyte solutions~\cite{Uematsu2018}.

Besides that, we focused our efforts on determining the interfacial boundary condition associated with the base toroidal flow state. We evidenced that the latter displays a `nearly solid' interface which can be described with a `quasi no\,--\,slip' boundary condition for the fluid velocity. A liquid\,--\,gas interface is classically modeled as a free surface, namely a no stress boundary. Conversely, a solid wall imposes such a strong constraint that the fluid velocity is zero everywhere on its surface. Here, however, we are in an intermediate situation as we assume that the water\,--\,air interface is partially covered with surfactants. 

Before proceeding further, let us precise what being `partially covered' means in our experimental conditions. According to a recent theoretical study~\cite{ThomasBickel2019}, a natural relaxation time for the elastic retraction of the surfactant film (Sec.~\ref{ResponseToLaserShutdowns}) is given by $\tau=\eta r/E_0$ ($\eta\sim 10^{-3}\,\mathrm{Pa}.\mathrm{s}$, water dynamic viscosity under standard conditions; $E_0$, Gibbs elasticity; $r$, radial distance from the heat source). Note that the above expression is based on dimensional analysis. Fig.\,\ref{fig:Fig8} yields a relaxation time $\tau\approx 4\,\mathrm{s}$ starting from a radial distance $r=2\,\mathrm{mm}$. Using then the relation $E_0=\textGamma_0 k_B T$ between the equilibrium Gibbs elasticity $E_0$ and surfactant concentration $\textGamma_0$, one ends up with the extremely small value $\textGamma_0\sim 100\,\,\text{molecules}/\mu\text{m}^2$~\cite{NoteOnEquilibriumGibbsElasticityExpressionInTheDiluteRegime}. The latter concentration is consistent with the very low surfactant contamination invoked either in~\cite{ThomasBickel2019} to explain the stiffening of the interface or in~\cite{HuAndLarson2005} to account for the suppression of Marangoni flows in evaporating droplets. Here the take\,--\,home message is that a surfactant concentration as minute as $\textGamma_0\sim 100\,\,\text{molecules}/\mu\text{m}^2$ is yet sufficient to provide the water surface with finite elasticity, thereby transforming it into a `nearly solid' interface that imposes a `quasi no\,--\,slip' boundary condition for the fluid velocity.\\

This fact is supported by the following observations:

\begin{itemize} 

\item As a first experimental proof, surface velocities are lower than subsurface ones in the toroidal regime (see Figs~\ref{fig:Fig5} and~\ref{fig:Fig6}). Such a `reversed stratification' of the velocities is at odds with the image of a free surface solely subject to a thermally\,--\,driven Marangoni flow, for which we naturally expect the velocity to be highest at the locus of flow onset, namely at the interface ($z=0$). 

\item The velocity scale of the pure thermocapillary flow given by~\cite{Wurger2014,ThomasBickel2019solo} $U=\gamma_T\textDelta T/2\eta\approx\gamma_T\mathcal{P}/4\pi\kappa\eta a$, where by definition $\gamma_T\,\dot{=}\,|\partial\gamma/\partial T|$, is on the order of $U\sim 1\,\mathrm{m.s^{-1}}$ (taking $\gamma_T\approx 0.144\,\,\mathrm{mN.\,m^{-1}.\,K^{-1}}$, $\mathcal{P}\sim 10\,\mathrm{mW}$, $a\sim 100~\mu$m, $\eta\sim 10^{-3}\,\mathrm{Pa}.\mathrm{s}$ and $\kappa\approx 0.6\,\,\mathrm{W}.\,\mathrm{m}^{-1}.\,\mathrm{K}^{-1}$). The latter theoretical order of magnitude exceeds by far the surface flow velocities (a few tens of $\mu$m/s) measured experimentally in the toroidal flow state. Note that some very recent studies (e.g.~\cite{Shmyrov2019}) also report a significant velocity drop at the interface relative to what is expected for a pure thermocapillary flow, with surface flow velocities that are one to two orders of magnitude lower in the case of a surfactant\,--\,laden interface.

\item The scaling law $r\sim t^{1/3}$ found out while tracking surface tracers (Fig.\,\ref{fig:Fig7}) departs from the theoretical prediction for a pure thermocapillary flow. The thermal P\'eclet number defined as $\mathrm{Pe}_{th}\,\dot{=}\,\,Ua/D$ (bead radius $a\sim 100~\mu$m; experimental Marangoni speed $U_\mathrm{exp}\sim 100~\mu$m/s; coefficient of thermal diffusivity $D\sim 10^{-7}\,\mathrm{m}^2.\,\mathrm{s}^{-1}$) is on the order of $10^{-1}$ in our operating conditions, meaning that heat transport is diffusion\,--\,dominated. Solving the heat diffusion equation in the steady regime yields the radial velocity~\cite{Wurger2014}
\begin{equation}
v_r(r,z=0)=U\left(\displaystyle\frac{a}{r}\right).
\end{equation}
The latter expression is inconsistent with $r\sim t^{1/3}$, the empirical scaling which corresponds to a faster power decay $v_r\,\dot{=}\,\mathrm{d}r/\mathrm{d}t\sim 1/r^2$. This discrepancy comes as one more hint supporting the existence of a surfactant elastic layer adsorbed at the water\,--\,air interface and responsible for the damping of the surface dynamics.
    
\end{itemize}    

\vskip0.1cm

The situation gets more involved when the forced flow is strong enough to overcome the elastic resistance of the surfactant\,--\,laden interface and create multipolar flows, in which case marked differences in the magnitude of the surface velocity are measured along the interface\,: intense flows are observed within the `channels', in contrast to what occurs in the regions close to vortex centers where the flow almost vanishes (compare Fig.~\ref{fig:Fig10} and Table~\ref{fig:Table1}). Most importantly, the channels are in a `passing state', i.e. characterized by velocities larger at the surface than in the bulk. The reverse trend, reminiscent of the `nearly solid' interface reported for the base torus (Sec.~\ref{ANearlySolidInterface}), is observed in the swirling flow regions. 

From a theoretical perspective, the main difficulty of the problem lies in its highly nonlinear nature. Indeed, the flow velocity, the temperature and the surfactant concentration fields are tightly coupled to one another through advection. In its highest level of generality, this problem calls for the joint solving of the Navier\,--\,Stokes equation together with the heat and mass advection\,--\,diffusion equations, a situation far too intricate to be addressed analytically. 

Still, inertia is irrelevant in the present case since we are working in the creeping flow regime\,: the Reynolds number defined as $\mathrm{Re}\,\dot{=}\,Ua/\nu$, with $U$ and $a$ the same typical velocity and length scales as above and $\nu$ the kinematic viscosity of water ($\nu\sim 10^{-6}\,\mathrm{m^2}.\,\mathrm{s}^{-1}$ under standard conditions), is on the order of $10^{-2}$ in usual experiments. Note that this goes against previous works which attribute the origin of the azimuthal instability to inertia. For instance, the authors of~\cite{ShternAndHussain1993} do predict the onset of the azimuthal instability but for a critical Reynolds number~\cite{NoteOnShternAndHussainNonStandardDefinitionOfTheReynoldsNumber} $\mathrm{Re}_c=115$ (see~\cite{ShternAndHussain1993}, Fig.\,8, limit of infinite Prandtl $\mathrm{Pr}$ number) which is much higher than the one encountered in our experiments. Based on this argument, we believe that the Shtern\,--\,Hussain scenario of the azimuthal instability presented in~\cite{ShternAndHussain1993} can be ruled out in our practical conditions. 

In the present situation, further simplification stems from the fact that the solutal P\'eclet number $\mathrm{Pe}_s$ is much larger than the thermal P\'eclet number $\mathrm{Pe}_{th}$, owing to a mass diffusivity $D_S\sim 10^{-10}-10^{-9}\,\mathrm{m^2}.\,\mathrm{s}^{-1}$ that is two to three orders of magnitude smaller than the coefficient of thermal diffusivity $D_T\sim 10^{-7}\,\mathrm{m^2}.\,\mathrm{s}^{-1}$. The physics of the system is hence dominated by the advective transport of surfactant molecules along the interface.

\begin{figure}
\centering
\includegraphics[width=0.8\columnwidth]{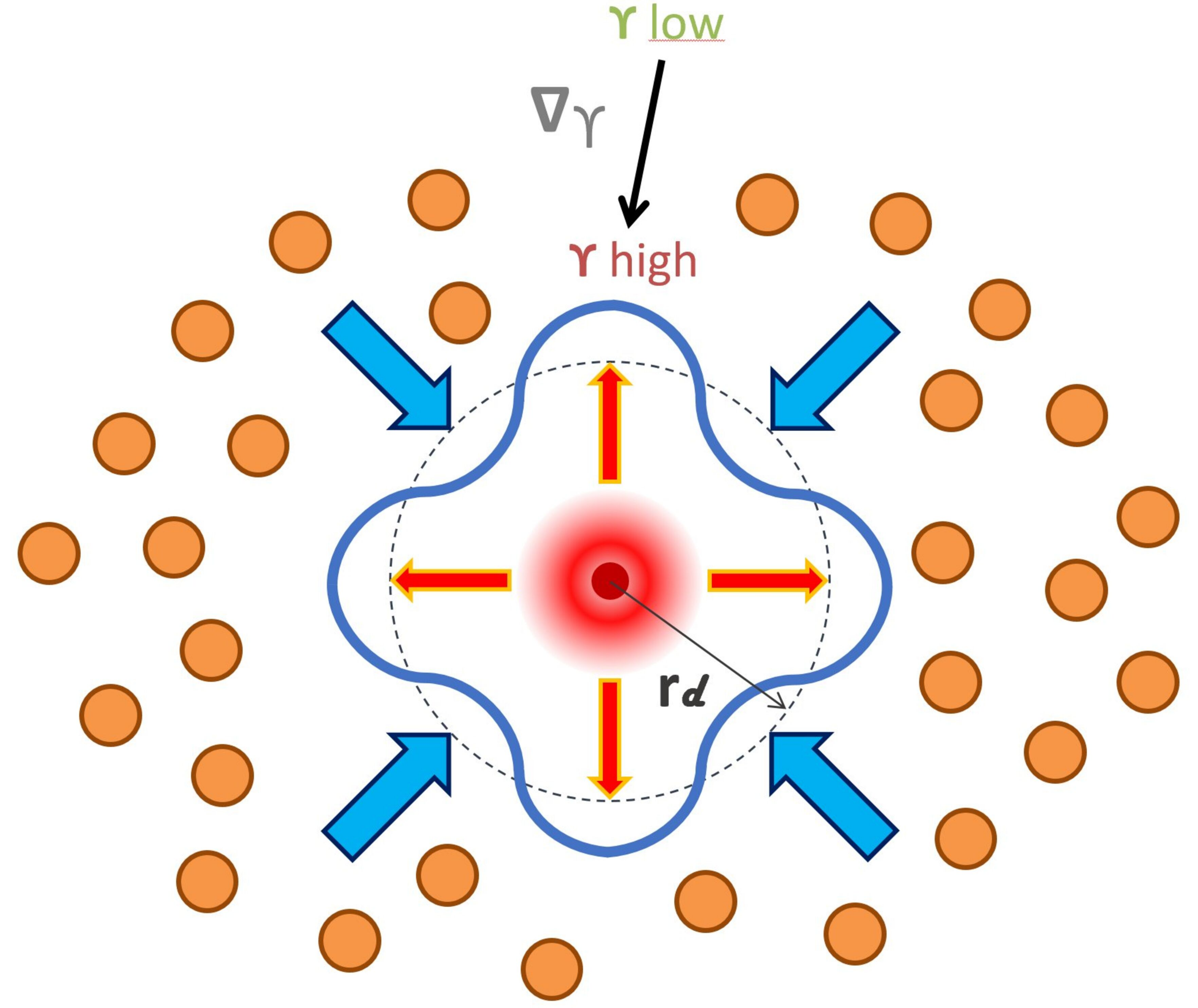}
\caption{Conjectural onset mechanism of the instability. Surfactant molecules (orange balls) are swept away beyond a radius $r_d$ from the heat source (red disk) and accumulate near the cell walls (not drawn) where they lower the surface tension ($\gamma_\mathrm{low}$). The competition between the centrifugal forced flow (red arrows) and the solutocapillary counterflow (blue arrows) deforms periodically the depletion elastic front (four\,--\,lobed line), eventually resulting in the azimuthal instability.}
\label{fig:Fig11} 
\end{figure}

The present study leads us to postulate a scenario for the onset of the azimuthal instability. Surfactants are swept by the dilatational flow towards the edges of the container, which induces a solutocapillary counterflow in response to the inhomogeneous distribution of surface\,--\,active impurities along the interface~\cite{Shmyrov2019}. We conjecture that the instability results from the elastic deformation of the depletion front, as illustrated in Fig.\,\ref{fig:Fig11}.

In fact, such an instability mechanism has been already proposed by Couder \textit{et al.} to account for multipolar flow patterns arising on the surface of soap films blown by a vertical air jet~\cite{Couder1989}. The scenario depicted here to explain the onset of the instability does not depend intrinsically on the nature\,---\,thermal, chemical, mechanical\,---\,of the source. The elasticity imparted to the interface by the surfactant film appears as a necessary condition for the onset of the instability. By the way, the fact that surface\,--\,active species destabilise the system goes against former studies in which they are shown to exert a stabilizing influence over convective instabilities~\cite{Berg1965}.     

To finish, let us come back to our starting point. The reported experiments, with a \textit{fixed} hot bead, tell us about the source mechanism for the orbital motion of the \textit{free} laser\,--\,heated particle (Fig.\,\ref{fig:Fig1}). The phenomenon clearly originates from the kind of azimuthal instability studied in the present work. As described by Girot \textit{et al.} in~\cite{Girot2016}, the particle is initially centered on the laser beam axis (black cross in Fig.\,\ref{fig:Fig1}), but escapes to a finite distance and starts orbiting around the axis. The configuration of a free hot sphere within a dipolar or quadrupolar flow pattern is presumably unstable. It is not surprising then that the hot particle moves out of the optical trap. The configuration sketched in Fig.\,\ref{fig:Fig1} cannot be static, which means that the pair of counter\,--\,rotating vortices acts as a propulsor, thereby transforming the hot particle into a microswimmer. What is still not understood is why the particle remains at a given distance from the laser beam axis. The latter property is far from evident. Hopefully, it could be elucidated based on a theory of the azimuthal instability that has not been developed yet.

Surface flow velocities on the order of $1\,\mathrm{mm.s^{-1}}$ are here attained with just a few tens of $\mathrm{mW}$ of laser power. So if the hot bead (radius $a\sim 100~\mu$m) were freed, it would be entrained by the self\,--\,induced flow at a speed about ten times its size per second. Thermocapillarity therefore appears as an efficient mechanism to achieve high\,--\,speed self\,--\,propulsion of active colloids~\cite{Frumkin2019}\,--\,\cite{ZottlAndStark2016}. In that respect, our study follows the line of recent theoretical works such as~\cite{Wurger2014}. On top of that, the generation of vortex pairs highlighted in the present paper offers an alternative to the autonomous motion of Janus colloids confined at a water\,--\,air interface, whose propulsion is fueled by an asymmetric catalytic reaction taking place on one side of the particle~\cite{Wang2017, Wang2015}. Despite its apparent simplicity the system investigated here may still harbor a plethora of interesting phenomena.\\

\section*{Supplementary material}

See the supplementary material for a video showing the elastic retraction of the interface at laser shutdown.

\begin{acknowledgments}

LOMA and CRPP acknowledge financial support from IDEX-Bordeaux under ``\,Propulsion de micro-nageurs par effet Marangoni\,'' PEPS program. We are grateful to B. Gorin, A. Mombereau, A. Girot and N. Dann\'e for fruitful collaboration, and to F. Nadal for stimulating discussions. We thank H. Kellay and C. Prad\`ere for the kind loan of a FLIR infrared camera (see note~\cite{NoteOnTracers}). The authors declare no conflict of interest.\\

\end{acknowledgments}

\section*{Data Availability}

The data that support the findings of this study are available from the corresponding author upon reasonable request.


\begin{thebibliography}{1}
\bibitem{NoteOnTracers}
We observed infrared images of the WA interface using a thermal camera (FLIR SC7000). IR images do not show the flow patterns but the temperature ($T$) map on the interface. As the $T$\,--\,distribution is connected to flow patterns through heat advection, symmetry breaking is readily revealed by IR images. Main features of $T$\,--\,maps turned out similar with and without polystyrene beads.
\bibitem{NoteOnLaserSheetsPositions}
Strictly speaking, we do not observe the flow at the WA interface but in a subsurface layer extending down to a few tenths of a millimeter below the surface. Indeed, the (low) deformation of the meniscus strongly impairs any proper observation of the interface based on laser sheet tomography. The vertical laser sheet is kept tangent to the bead as the obstruction by the optic fiber located on the cell centerline prevents us from performing diametral sections of the flows.
\bibitem{NoteOnPIVInaccuracyOnHorizontalCuts}
In principle, it is possible to get interfacial velocity maps by running PIV on horizontal cuts. In practice, however, large errors may be committed below the onset of the instability due to surface and subsurface tracers moving at markedly different speeds over the whole thickness of the laser sheet (Sec.~\ref{BaseFlowState}). The latter issue disappears in case of perfect slip at the interface for which surface and subsurface tracers move at comparable speeds (Sec.~\ref{QuadrupolarFlow}).
\bibitem{NoteOnImperfectAxisymmetry}
As can be noticed in the upper right corner of the top figure, the axisymmetry of the base flow is not perfect\,: the streamlines are not strictly radial but slightly curved along some preferred direction. The latter varies from one experiment to another leaving the impression that it is due to parasitic temperature gradients and/or geometric flaws of the system (e.g., the rough surface of the bead).
\bibitem{NoteOnTrajectoryReconstruction}
In practice, there are at least two reasons why surface tracers can be tracked only over short distances relative to the cell width\,: the vertical laser sheet is not perfectly aligned with the radial trajectories of the tracers~\cite{NoteOnLaserSheetsPositions} and, to a lesser extent, thermal agitation tends to push tracers out of the viewing plane.
\end{thebibliography}
\end{document}